\documentclass{iau}
\usepackage{graphicx,natbib}
 
\newcommand{\apj}{ApJ}           
\newcommand{\mnras}{MNRAS}       

\newcommand{\aap}{A\&A}

\newcommand{\aj}{AJ}

\newcommand{\HI}{H\,{\sc i}}
\newcommand{\Ha}{H$\alpha$}
\newcommand{\degr}{^\circ}
\newcommand{\kms}{~km\,s$^{-1}$}

\newcommand{\MHI}{M$_{\rm HI}$}
\newcommand{\Msun}{~M$_\odot$}

\title[The Local Universe: Galaxies in 3D]
      {The Local Universe: Galaxies in 3D}

\author[B\"arbel S. Koribalski]
       {B\"arbel S. Koribalski}

\affiliation{CSIRO Astronomy and Space Science, Australia Telescope 
   National Facility,\\ P.O. Box 76, Epping, NSW 1710, Australia \\
email: {\tt Baerbel.Koribalski@csiro.au}}

\pubyear{2014}
\volume{309}
\jname{Galaxies in 3D across the Universe}
\editors{B. L. Ziegler, F. Combes, H. Dannerbauer, M. Verdugo, eds.}

\begin{document}

\maketitle

\begin{abstract}
Here I present results from individual galaxy studies and galaxy surveys 
in the Local Universe with particular emphasis on the spatially resolved 
properties of neutral hydrogen gas. The 3D nature of the data allows 
detailed studies of the galaxy morphology and kinematics, their relation to
local and global star formation as well as galaxy environments. I use new 
3D visualisation tools to present multi-wavelength data, aided by 
tilted-ring models of the warped galaxy disks. Many of the algorithms and 
tools currently under development are essential for the exploration of 
upcoming large survey data, but are also highly beneficial for the analysis 
of current galaxy surveys.

\end{abstract}

\firstsection

\section{Introduction}

Analysing the well-resolved stellar and gas kinematics of disk galaxies 
provides insights into their rotational and non-rotational components, both of 
which can be measured as a function of radius and disk height. The resulting 
rotation curves reflect the overall mass distribution of galaxies (eg., Bosma 
2004), including their very large dark matter halos, which extend well beyond 
the observed disks (e.g., Jones et al. 
1999; Warren et al. 2004; Kreckel et al. 2011; Koribalski \& Lopez-Sanchez 
2009; Westmeier et al. 2011, 2013).  High-resolution spectroscopic data cubes 
(targeting, eg., the \HI, CO, and \Ha\ spectral lines) also allow us to 
determine the 3D shape of galaxies, including their warped disks. The various 
shapes of 3D models derived with TiRiFiC, the {\em Tilted Ring Fitting Code}, 
are nicely illustrated by J\'ozsa (2007). I then construct 3D visualisations 
of the observed galaxies based on multi-wavelength imaging and spectral line 
cubes at various angular resolutions, allowing me to analyse and improve the 
3D representation of each galaxy until it reflects the observed data and 
knowledge derived from the data. Another benefit of the re-constructed 3D 
particle model is the ability to view each galaxy from any angle and produce
fly-through movies (as presented for the galaxy M\,83). In this first attempt, 
I used high-resolution optical and ultraviolet images as well as \HI\ spectral 
line cubes from the Australia Telescope Compact Array (ATCA). \\ 

Optical galaxy surveys with ever more powerful integral fields units (IFUs), 
such as ATLAS-3D (Cappellari et al. 2011), CALIFA (S\'anchez et al. 2012) and 
SAMI (Allen et al. 2014), are now delivering high-resolution data cubes. Radio 
synthesis telescopes like the ATCA, which consists of six 
22-m dishes, have been recording large spectral line data cubes for over two 
decades. The ATCA primary beam is $\sim$0.5 degr at 21-cm; \HI\ emission from 
nearby galaxies within this large field of view can be mapped with angular 
resolutions of up to $\sim$10 arcsec and velocity resolution of less than 
1\kms. Much higher resolutions are typically used when observing molecular 
lines (ATCA receivers cover frequencies from 1 to 105~GHz observable with
two 2~GHz-wide bands; Wilson et al. 2011). \\

\begin{figure}
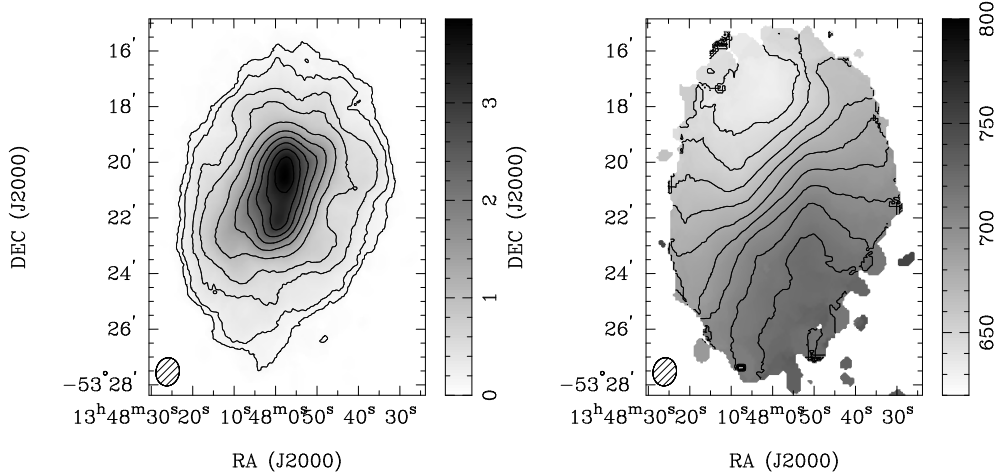
 
\centering
 \includegraphics[width=6.5cm]{j1348-53.mom0.ps}
 \includegraphics[width=6.5cm]{j1348-53.mom1.ps}
\caption{ATCA \HI\ moment maps of the low-surface brightness galaxy 
   ESO174-G?001 obtained as part of the {\em Local Volume HI Survey} (LVHIS) 
   project (Koribalski 2008). The \HI\ distribution (left) extends far beyond 
   the stellar disk, while the mean \HI\ velocity field (right) shows a 
   regular, but rather twisted rotation pattern.}
\label{fig:koribalski-fig1}
\end{figure}

ASKAP, the Australian Square Kilometre Array Pathfinder (Johnston et al. 
2008), which will be fully equipped with novel wide-field phased array feeds 
(PAFs), is just starting to deliver the first well-resolved \HI\ maps of 
nearby galaxies and groups. We are currently observing with six of the 36 
antennas, forming nine beams (each with FWHM $\sim$1~degr) which can be placed 
anywhere within the 30 sq degr field-of-view. After successful ASKAP \HI\ 
mapping the Sculptor Group galaxies NGC~253 and NGC~247 plus the area in 
between this wide pair, we are now imaging the nearby gas-rich galaxy group 
IC\,1459 and its environment (Serra et al. 2015). The recording bandwidth 
is 300~MHz divided into around 17\,000 channels giving a velocity resolution 
of 4\kms. The angular resolution of the current 6-antenna array, known as 
the Boolardy Engineering Test Array (BETA; Hotan et al. 2014) is $\sim$1 
arcmin; the frequency coverage is 700 to 1800~MHz. \\

ASKAP is a powerful 21-cm survey machine and will --- once fully equipped with 
the equivalent of wide-field IFUs --- be mapping the neutral hydrogen line 
in emission and absorption over the whole southern sky and a good fraction of
the northern sky. Several large ASKAP \HI\ surveys are planned. Here I briefly 
introduce WALLABY, a 21-cm survey of the sky ($\delta < +30\degr$; $z < 0.26$) 
which in about one year of observing time will detect more than 500\,000 
galaxies in the \HI\ spectral line (Duffy et al. 2012, Koribalski 2012b), ie. 
a factor $\sim$20 more than currently catalogued. Novel Chequerboard PAFs are 
providing 30 sq degr instantaneous field-of-view. In WALLABY $\sim$1000 
galaxies will have \HI\ diameters larger than 5~arcmin ($>$10 beams) and 
$\sim$5000 galaxies will have major-axis \HI\ diameters greater than 2.5 
arcmin ($>$5 beams), allowing us to study in detail their morphology, 
kinematics and mass distribution. The number would rise to $1.6 \times 10^5$ 
galaxies if all 36 ASKAP antennas could be used; the additional six antennas 
provide baselines up to 6~km, resulting in an angular resolution of 10$''$. 
Creating highly reliable and complete source catalogs requires sophisticated 
source-finding algorithms as well as accurate source parametrisation. We are
aiming to achieve this with our new {\em Source Finding Application} (SoFiA;
Serra et al. 2014) and \HI\ profile fitting with the versatile "Busy Function" 
(Westmeier et al. 2014). For an overview on continuum and spectral line source 
finding see the {\em PASA Special Issue} (Koribalski 2012a). \\ 

\section{Kinematic Modelling} 
A common way of analysing a galaxy's \HI\ velocity field is to carefully fit 
tilted ring models (de Blok et al. 2008, Oh et al. 2011). The resulting
residual velocity field is a good indicator of the goodness of fit and any
remaining peculiar (non-rotational) structures. Tilted ring fits retrieve 
structural parameters, such as position angle, inclination and rotational 
velocity, as a function of radius when a galaxy is moderately well resolved. 
Several well-tested algorithms/programs exist, most of which are applied to 
the 2D velocity field. In some cases, for example edge-on disks and large 
well-resolved \HI\ disks of galaxies, 3D modelling of the \HI\ data cubes is
of advantage (eg., Kamphuis et al. 2013). 

\begin{figure}[h] 
\centering
\includegraphics[width=0.7\columnwidth]{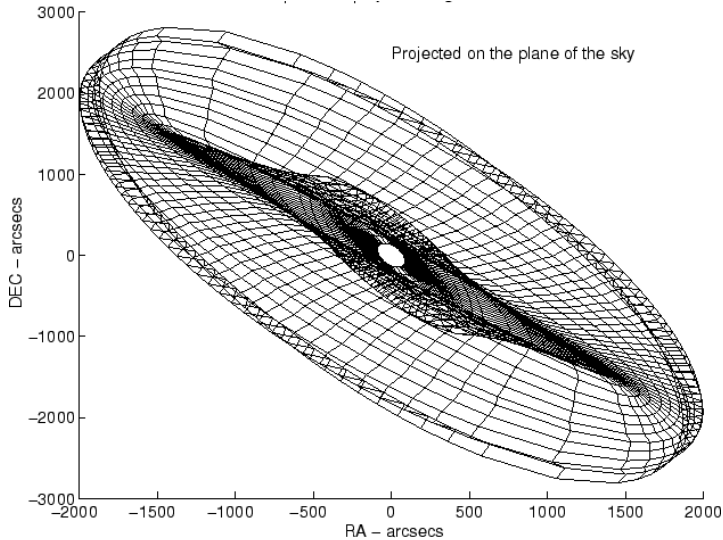} 
\caption{Example 3D model of a warped spiral galaxy.}
\label{fig:koribalski-fig2}
\end{figure}

The WALLABY kinematics working group (led by Kristine Spekkens) has started 
to create a modular package that allows to input spectral data cubes of 
individual galaxies, calculate \HI\ velocity fields (eg., as shown in 
Fig.~\ref{fig:koribalski-fig1}) and fit them with one or more of the 
algorithms described below. We base this approach on our successful SoFiA 
package (Serra et al. 2014). Assuming circular symmetry, the fitted rings 
allow an accurate description of warped galaxy disks where both inclination, 
$i(r)$, and position angle, $PA(r)$, vary with radius $r$. This is necessary 
to correctly deproject the measured rotational velocities to derive the 
galaxy rotation curve, $v_{\rm rot}^{i}(r)$. Rogstad et al. (1974) 
introduced the tilted-ring model fitting, which is widely used through the 
{\em Rotcur} program and applied to 2D velocity fields (see, eg., 
Fig.~\ref{fig:koribalski-fig2}). {\em TiRiFiC} uses spectroscopic data 
cubes to obtain tilted ring model fits of galaxies; it is now available 
as a stand-alone routine. This approach is essential for galaxy disks seen 
edge-on and allows sophisticated modelling of the disk thickness and surface 
brightness distribution (Kamphuis et al. 2013, 2014). {\em DiskFit} was 
created for the kinematic modeling of barred spiral galaxies. It does not 
allow the fitting of warped disk where the inclination and position angles 
vary with radius. DiskFit builds on {\em Velfit} (Spekkens \& Sellwood 2007). 
The {\em Kinemetry} technique (Krajnovic et al. 2006) was devised to model 
the stellar velocity fields of early-type galaxies. 

\section{The Spiral Galaxy M\,83} 
In the following I will use the grand-design spiral galaxy as an example of
3D \HI\ mapping. The \HI\ envelope of M\,83 is much more extended than the
well-known stellar disk. A prominent tidal arm hints at gravitational
interactions
with its dwarf galaxy neighbours while the western edge of the gas appears
compressed, possibly affected by ram pressure stripping. The \HI\ velocity
field shows both the rotation and the changing orientation of the gaseous 
disk. By modelling the disk we are able to reconstruct the true 3D shape of
a galaxy and place it within the group volume. In Section~5 I give more 
details on the 3D visualisation of stars and gas using the ray-tracing 
software Splotch. Currently, we can model 
several hundred galaxies in this way. Future large-scale ASKAP \HI\ surveys 
will allow us to do this for several thousand galaxies, obtaining a true 3D
dynamic picture of the nearby Universe. \\

\begin{figure} 
\centering
 \includegraphics[width=13.5cm]{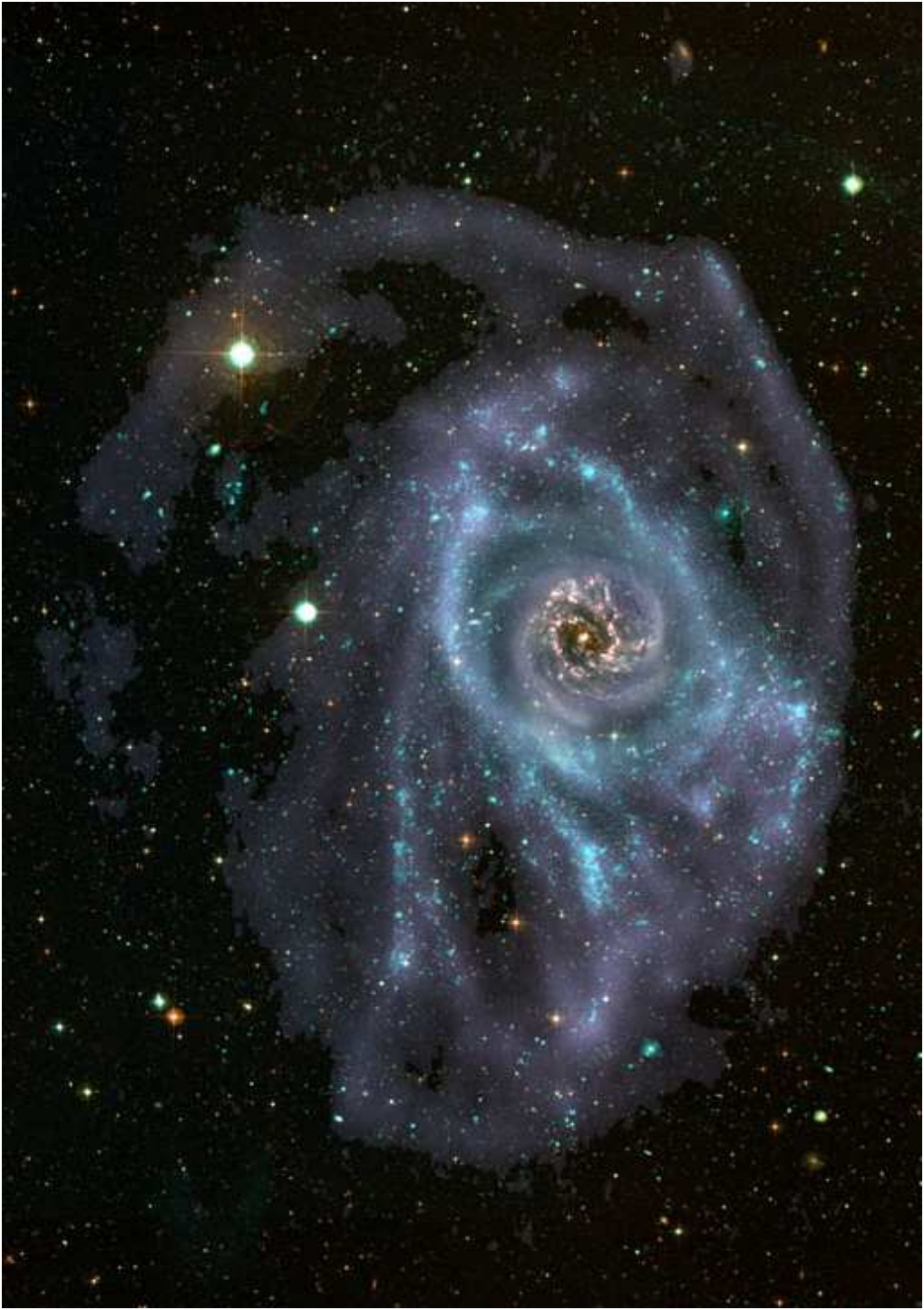} 
\caption{Multi-wavelength image of the spiral galaxy M\,83. The large neutral 
  hydrogen (\HI) disk, shown in blue, was mapped with the ATCA and extends
  well beyond the stellar disk. For the color composition we also used GALEX 
  NUV+FUV (light blue), DSS $R$-band (green), and 2MASS $J$-band (red) images.
  (Image credit: \'Angel R. L\'opez-S\'anchez).}
\label{fig:koribalski-fig3}
\end{figure}

Our large-scale \HI\ mosaic of the beautiful, grand design spiral galaxy
M\,83 (NGC~5236, HIPASS J1337--29), obtained by combining ATCA interferometric 
and Parkes single-dish data (see Fig.~\ref{fig:koribalski-fig3}), reveals an 
enormous \HI\ disk with 
a diameter of $\sim$80 kpc, several times larger than M\,83's optical Holmberg 
diameter. Here we adopt a distance of $D$ = 4.5~Mpc. While the inner disk of 
M\,83 rotates remarkably regular, the \HI\ gas dynamics appear increasingly 
peculiar towards the outer regions which show clear signs of tidal disruption. 
The most prominent tidal features of M\,83 are the one-sided outer \HI\ arm 
which can be traced over 180 degr from the western to the eastern side, and
the spectacular stellar stream, consisting of mainly old stars, to the north; 
their origin and possible relation are explored. M\,83 is surrounded by 
numerous dwarf galaxies and, given its dynamical mass of about $5 \times 
10^{11}$\Msun, is likely to attract and accrete them in regular intervals. \\

From a very deep optical image of M\,83 by Malin \& Hadley (1997) we estimate
a stellar diameter of $\sim$22$' \times 19'$ (ie, 29~kpc $\times$ 25~kpc) for 
the disk at a $B$-band surface brightness of $\sim$28 mag\,arcsec$^{-2}$. For 
comparison, the optical $B_{25}$ diameter of M\,83 is $12.9' \times 11.5'$ 
(de Vaucouleurs et al. 1991). 
Beyond the faint stellar disk, the deep optical image reveals three peculiar 
features: a prominent north-western {\em stellar stream} at a distance of 
$\sim$18$'$ from the center of M\,83, a {\em small arc} towards the 
north-east at a distance of $\sim$11$'$, and a {\em southern ridge}.
GALEX FUV emission is clearly detected 
in the {\em southern ridge} as well as the western part of the {\em small 
arc} which we identified in the deep optical and \HI\ images of M\,83
(see Fig.~\ref{fig:koribalski-fig3}).
The mean $B$-band surface brightness of the extended stellar stream to the 
north of M\,83 is around 27 mag\,arcsec$^{-2}$. It covers an area of about 
10 arcmin$^2$ or about 17 kpc$^2$. 
The \HI\ distribution (0. moment) of M\,83 is most remarkable. No longer does 
this grand-design spiral look regular and undisturbed. The \HI\ maps show 
streamers, irregular enhancements, an asymmetric tidal arm, diffuse emission, 
and a highly twisted velocity field, much in contrast to its regular appearance
in short-exposure optical images. M\,83's \HI\ mass (\MHI\ = 7.8 
($\pm$0.5) $\times 10^9$\Msun) is more than twice that of an M$_{\rm HI}^*$ 
galaxy. The \HI\ appearance of M\,83 clearly 
suggests that it has been and possibly still is interacting with neighbouring 
dwarf galaxies. The effect of this interaction on the dwarfs can of course
be rather devastating; it is quite likely that M\,83 has accreted dwarf 
galaxies in the past as the stellar and gaseous streams/tails. 
The eastern-most \HI\ emission of M\,83 which forms part of its peculiar, outer
arm lies $\sim34.5'$ or 45 kpc away from the center of M\,83. And the 
the dwarf irregular galaxy NGC~5264 lies at a projected
distance of only $25.5'$ or 33 kpc from the eastern \HI\ edge of M\,83. 
For NGC~5264 we measure \MHI\ = $6 \times 10^7$\Msun\
and calculate a total dynamical mass of $\sim1.2 \times 10^9$\Msun.
We also detected the dwarf irregular galaxy UGCA\,365 to the north of M\,83, 
just outside its large \HI\ envelope. We measure \MHI\ = $1.5 \times 
10^7$\Msun\ and a total dynamical mass of $\sim1.2 \times 10^9$\Msun. --- 
For comparison, the \HI\ disk of the Circinus galaxy has a diameter of 
$\sim80'$ or $\sim$100 kpc at an adopted distance of 4.2 Mpc (For, Koribalski
\& Jarrett 2012), $\sim5\times$ larger than the extrapolated Holmberg diameter 
of $\sim17'$ (Freeman et al. 1977).

\begin{figure} 
\centering
 \includegraphics[width=9.6cm]{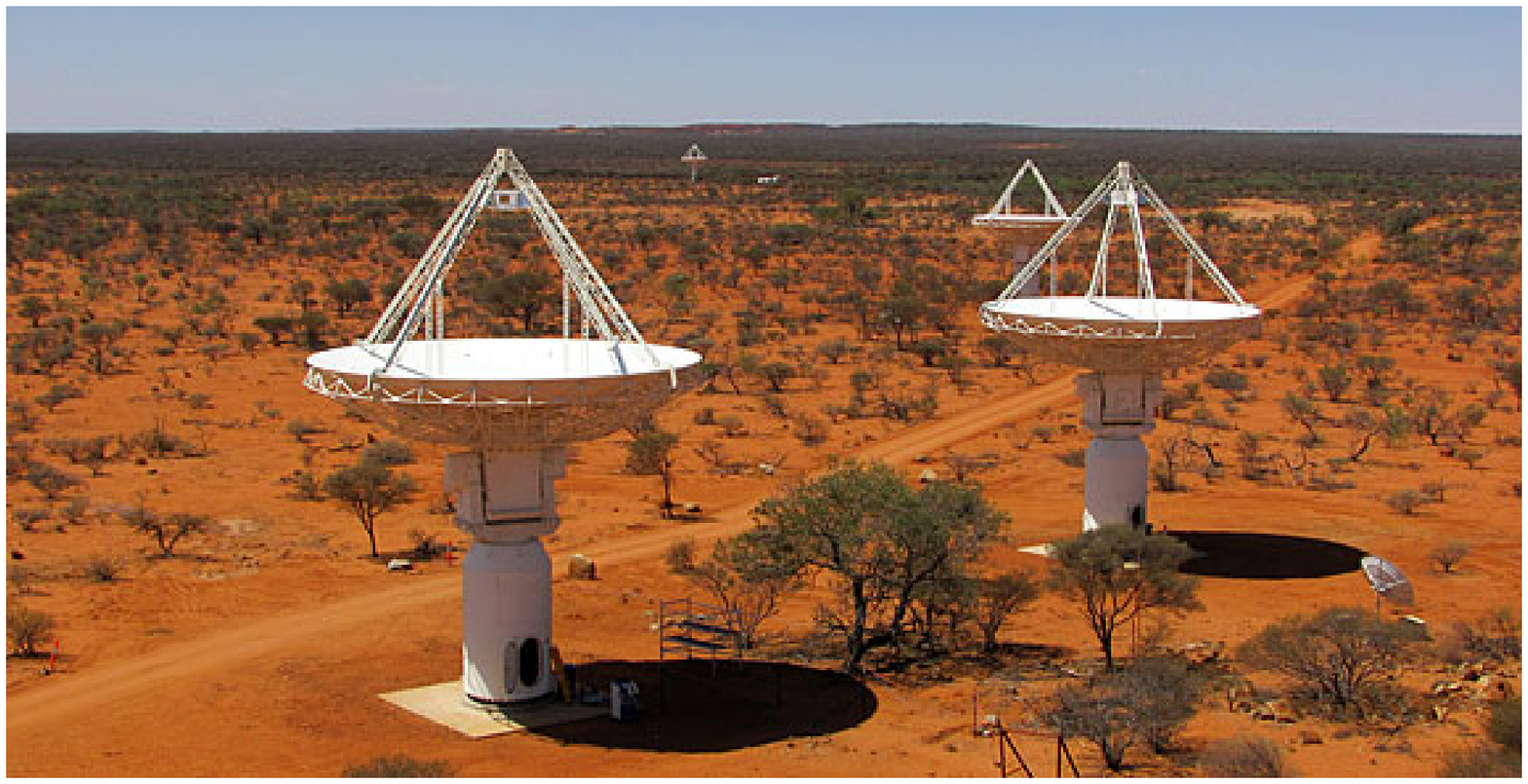}
 \includegraphics[width=3.4cm]{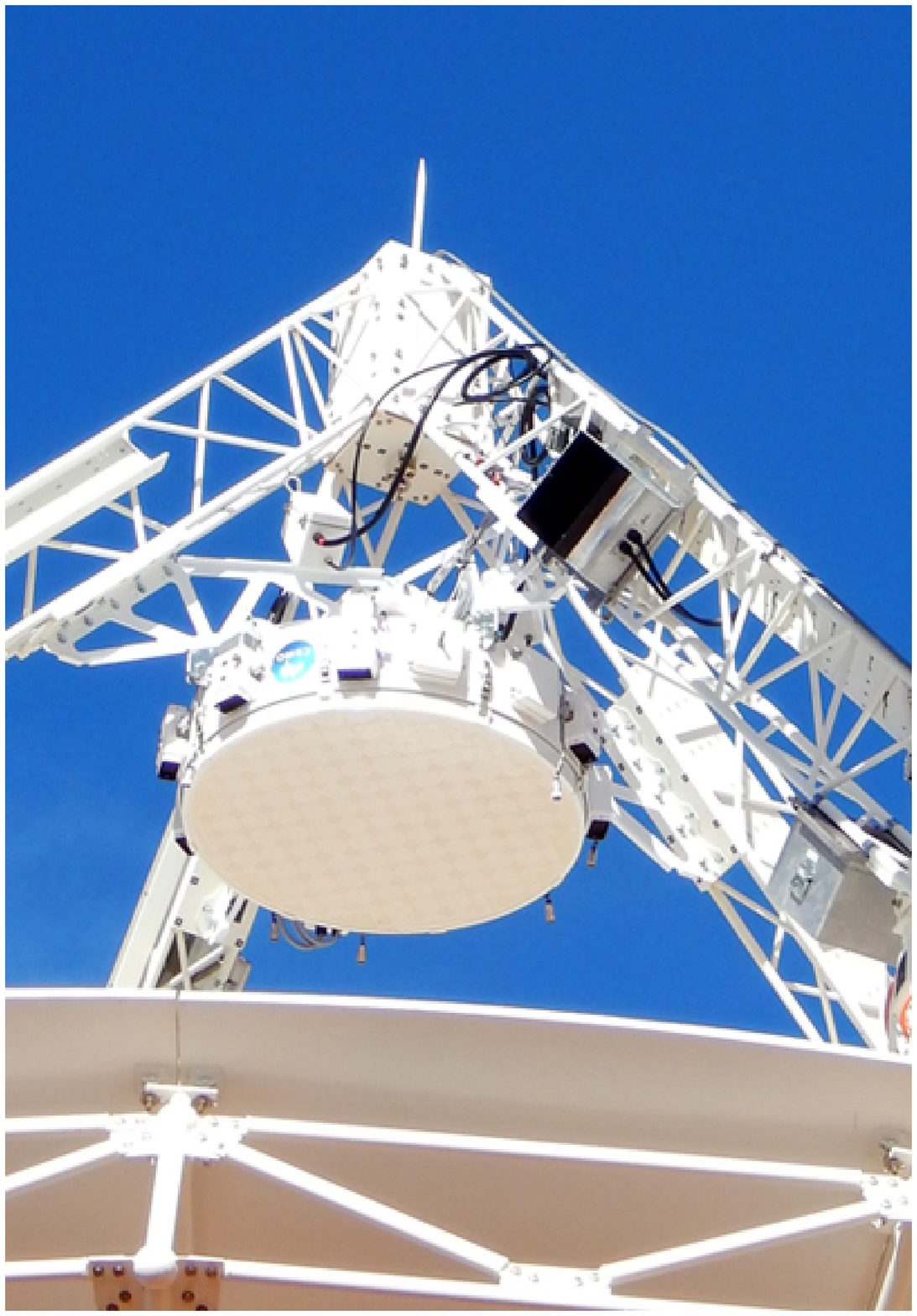}
\caption{--- (Left) Three of the 36 ASKAP antennas at the Murchison Radio 
  astronomy observatory (MRO) in Western Australia. --- (Right) The first 
  Mk\,II PAF was installed on ASKAP antenna 29 in September 2014. (Image 
  credits: Ant Schinckel, CSIRO).}
\label{fig:koribalski-fig4}
\end{figure}

\section{SKA Pathfinder HI Surveys} 

We have come a long way since the discovery of the 21-cm spectral line by Ewen 
\& Purcell in 1951. Nevertheless, detecting \HI\ emission in a Milky Way-like 
galaxy at redshift $z = 1$ will require the Square Kilometre Array (SKA; 
Rawlings 2011; Obreschkow et al. 2011). Several SKA pathfinder and precursor 
telescopes are currently testing new technologies and are used to develop 
reliable data processing pipelines. Much of the data is too large to store 
long-term, so data processing, quality control, imaging and source finding 
has to be done as the raw data is being recorded. \\

Several tens of thousands galaxies have so far been detected in the 21-cm 
\HI\ line, the vast majority with single dish radio telescopes. The intrinsic 
faintness of the electron 
spin-flip transition of neutral atomic hydrogen (rest frequency 1.42~GHz) 
makes it difficult to detect \HI\ emission from individual galaxies at large 
distances. To study the \HI\ content of galaxies and diffuse \HI\ filaments 
between galaxies, we need radio synthesis telescopes with large collecting 
areas, low-noise receivers and large fields of view. \\

ASKAP consists of
$36 \times 12$-m antennas and is located in the Murchison Shire of Western
Australia (see Fig.~\ref{fig:koribalski-fig4}). Of the 36 antennas, 30 are 
located within a circle of $\sim$2~km diameter, while six antennas are at 
larger distances providing baselines up to 6~km. Six ASKAP dishes are 
currently equipped with first-generation (Mk\,I) Chequerboard PAFs. The 
instantaneous field-of-view of the ASKAP 
PAFs is 5.5 deg $\times$ 5.5 deg, ie. 30 square degrees, making ASKAP a 21-cm 
survey machine. The WSRT APERTIF upgrade employs Vivaldi PAFs, delivering a 
field-of-view of 8 square degrees (Verheijen et al. 2008). \\

{\bf WALLABY}, the {\em Widefield ASKAP L-band Legacy All-sky Blind surveY}
(led by me and Lister Staveley-Smith; see Koribalski et al. 2009), will cover 
75\% of the sky ($-90\degr < \delta < +30\degr$) over a frequency range from 
1.13 to 1.43~GHz (corresponding to $-2000 < cz < 77,000$\kms) at resolutions 
of $30''$ and 4\kms. WALLABY will be carried out using the inner 30 antennas 
of ASKAP, which provide excellent $uv$-coverage and baselines up to 2~km. 
High-resolution ($10''$) ASKAP \HI\ observations using the full 36-antenna 
array will require further computing upgrades. {\bf WNSHS}, the {\em Westerbork
Northern Sky HI Survey} (led by Guyla J\'ozsa), will cover a large fraction 
of the northern sky ($\delta > +27\degr$) with APERTIF over the same frequency 
range as WALLABY with ASKAP. Both \HI\ surveys combined will achieve a true 
all-sky survey with unprecedented resolution and depth. The science goals of 
both surveys are well developed and complement, as well as enhance, each other.
For a summary see Koribalski (2012b). 
{\bf WALLABY} and {\bf WNSHS} are made possible by the development of phased 
array feeds, delivering a much larger field-of-view than single feed horns or 
multi-beam systems. WALLABY will take approximately one year (ie 8 hours per 
pointing) and deliver an rms noise of 1.6 mJy\,beam$^{-1}$ per 4\kms\ channel. 
WALLABY is a precursor for future \HI\ surveys with SKA Phase\,I and II, 
exploring the role of atomic hydrogen in galaxy formation and evolution. 
Using the \HI-detected galaxies over two thirds of sky we will be able to 
unveil their large-scale structures and cosmological parameters. For 
nearby galaxies, we can detect their extended, low-surface brightness 
disks as well as gas streams and filaments between galaxies. 

\section{3D Visualisation of gas and stars in galaxies} 

Animations of several thousand HIPASS galaxies show their 3D distribution 
in the nearby Universe ($z < 0.03$), based on their measured positions, 
velocities / distances and \HI\ masses (created by Mark Calabretta and 
available on-line). Large-scale structures such as the Supergalactic 
Plane and the Local Void (see Koribalski et al. 2004) are clearly visible.
The resolution of HIPASS, the \HI\ Parkes All Sky Survey, is 15.5 arcmin
and 18\kms.  Once WALLABY data are in hand (resolution 30~arcsec and 
4\kms), we should be able to create animations of 500\,000 galaxies as well
as detailed 3D models of $\sim$5000 galaxies (ie, all HIPASS galaxies)
by a 3D multi-wavelength rendering of each respective galaxy. 
Multi-wavelength images and spectral line data cubes of galaxies allow us to
measure their stellar, gas and dark matter properties. Visualisation packages
such as {\sc KARMA} provide a range of tools to interactively view
2D and 3D data sets as well as apply mathematical operations. This allows
not only the quick inspection and evaluation of multiple images, spectra and 
cubes, but also the production of beautiful multi-color images and animations. 
To improve our understanding of galaxy formation and evolution, we need to 
include models and theoretical knowledge together with observations of galaxy
disks and halos. By fitting and modelling the observed gas distribution and 
kinematics of extended galaxy disks, we derive their 3D shapes and rotational 
velocities. Visualisation can then be employed to combine the actual data 
with our derived knowledge to re-construct the most likely 
3D representation of each galaxy. By adding time as the fourth dimension one 
can also visualise the evolution of galaxies and the Universe (e.g., see the 
{\em 4D~Universe} visualisation by Dolag et al. 2008). 
{\sc SPLOTCH} is a powerful and very flexible ray-tracer software tool
which supports the visualisation of large-scale cosmological simulation data 
(Dolag et al. 2008). It is publicly available and 
continues to be enhanced. A small team is currently working on supporting the 
visualisation of multi-frequency observational data to achieve realistic 3D 
views and fly-throughs of nearby galaxies and galaxy groups. 

\section*{Acknowledgements}

\noindent
I like to thank the conference organisers for inviting me to participate in
such a fantastic conference in the beautiful city of Vienna. Furthermore, I 
thank my collaborators, in particular Claudio Gheller and Klaus Dolag from 
the 3D visualisation team, my wonderful WALLABY team, and my colleagues 
Peter Kamphuis and Tiffany Day who continue to improve the \HI\ models of 
M\,83. I gratefully acknowledge financial support from the IAU.

\end{document}